# High-throughput optical neural networks based on temporal computing

Shuang Zheng[1,2,3,4,†], Jiawei Zhang[1,2,3,4,†], Weizhen Yu[1,2,3,4] & Weifeng Zhang[1,2,3,4,*]

[1]Radar Research Lab, School of Information and Electronics, Beijing Institute of Technology, Beijing 100081, China

[2]Key Laboratory of Electronic and Information Technology in Satellite Navigation (Beijing Institute of Technology), Ministry of Education, Beijing 100081, China

[3]Chongqing Innovation Centre, Beijing Institute of Technology, Chongqing 401120, China

[4]Chongqing Key Laboratory of Novel Civilian Radar, Chongqing 401120, China

[†]These authors contributed equally: Shuang Zheng, Jiawei Zhang.

[*]Correspondence and requests for materials should be addressed to Weifeng Zhang.

(email: weifeng.zhang@bit.edu.cn)

## Abstract

An emerging generative artificial intelligence (AI) based on neural networks starts to grow in popularity with a revolutionizing capability of creating new and original content. As giant generative models with millions to billions of parameters are developed, trained and maintained, a massive and energy-efficient computing power is highly required. However, conventional digital computers are struggling to keep up with the pace of the generative model improvements. In this paper, we propose and demonstrate high-throughput optical neural networks based on temporal computing. The core weighted summation operation is realized with the use of high-speed electro-optic modulation and low-speed balanced photodetection. The input data and weight are encoded in a time sequence separately and loaded on an optical signal via two electro-optic modulators sequentially. By precisely controlling the synchronization time of the data and weight loading, the matrix multiplication is performed. Followed by a balanced photodetector, the summation is conducted, thanks to the electron accumulation of the inherent electronic integrator circuit of the low-speed photodetector. Thus, the linear weighted summation operation is implemented based on temporal computing in the optical domain. With the proposed optical linear weighted summation, a fully-connected neural network and convolutional neural network are realized. Thanks to the high-speed feature of temporal computing, a high data throughput of the optical neural network is experimentally demonstrated, and the weighting coefficients can be specified on demand, which enables a strong programmability of the optical neural network. By leveraging wavelength multiplexing technology, a scalable optical neural network could be created with a massive computing power and strong reconfigurability, which holds great potential for future giant AI applications.

## Introduction

A neural network is a computing model featuring interconnected nodes or neurons in a layered structure for information processing that mimics a human brain [1]. By identifying the patterns and structures among existing data with the use of neural networks, an emerging generative artificial intelligence (AI) starts to grow in popularity with a revolutionizing capability of creating new and original content [2-6]. To handle more complex patterns, giant generative models with millions to billions of parameters are developed, trained and maintained. As a result, a massive and energy-efficient computing power is highly required on the computation infrastructure [7, 8]. Due to the unsustainability of Moore's law and the limitations of the von Neumann architecture, conventional digital computers are struggling to keep up [9, 10]. By leveraging the unique properties of photons including broad bandwidth, low latency, and high energy efficiency, analog optical computing is envisioned as a potential solution to serve as a computing accelerator [11, 12].

Recently, much efforts have been directed to the study of optical neural networks (ONNs), and different realizations have been proposed [13, 14]. In an ONN, a linear weighted summation operation or a vector–matrix multiplication operation is essential. To date, several experimental implementations of optical weighted summation have been reported based on free-space optics or integrated optics. In free-space optics, the spatial light intensity distributions in the input and output planes can be straightforwardly regarded as the input and output vectors. In the light propagation direction, multiple layers of cascaded diffractive optical elements (DOEs) or spatial light modulators (SLMs) are placed where the weight coefficient vectors can be encoded. After passing through the DOEs or SLMs, the light intensity on the plane can be collected with a lens or lens array as the weighted summation output [15-18]. Thanks to the pixel reconfigurability of the DOEs and SLMs, the input light intensity pattern and weighting coefficients can be flexibly tuned, which is of great benefit to the agility and adaptivity of the resulted ONNs. However, the key disadvantage is that the pixel number determines the size of the input light intensity pattern and weighting coefficients, which finally limits the scaling of the neural network.

Compared to free-space optics, integrated optics holds great potential for large-scale integration, ultra-small footprint, and highly-reduced energy consumption. Due to strong light confinement and compatibility with the CMOS technology, silicon photonic integrated circuits (PICs) have received an extensive study in the past decades, and silicon-based integrated ONN solutions have also been frequently proposed in recent years [19-30]. Typical on-chip ONNs configurations have a mesh structure of cascaded Mach–Zehnder interferometers (MZIs) or microring resonator-based wavelength division multiplexing (WDM) structure. The MZIs structure usually have a relatively large size in the scale of hundreds of micrometres, which leads to a limited number of the MZIs integrated on the same die. Consequently, it is very challenging to scale up an MZI-based ONN since the number of the MZIs is required to increase quadratically with the input data dimension for matrix multiplication. Currently, the reported largest MZI mesh network has an input vector size of 4×1 [20, 26, 29]. Unlike the MZI, a micro-ring resonator (MRR) has a much smaller size in the scale of tens of micrometres. Optical signals with different wavelengths are used as the input vector and undergo different spectral filtering of the MRRs as weighting coefficients multiplication [30]. Therefore, the vector size of the ONN depends on the wavelength number of the optical sources. In addition, because of temperature sensitivity, an electrical control circuit is needed to reduce resonance wavelength fluctuations of the MRRs, which inevitably increases the cost of the ONNs.

Although different neural networks either on free-space optics or on integrated optics have been demonstrated, their main problem is that the scaling of the ONNs is severely restricted by the number of the pixels, or MZIs or wavelength channels since the data or weight are encoded in a parallel way. Inspired by human brain information process based on the timing of events, temporal or time-based computing encodes the data and weight in a time sequence for multiplication and summation, which offers a promising manner to perform big-data processing and machine learning efficiently [31, 32]. Recently, a photonic accelerator based on temporal computing is proposed, in which both the inputs and weights are encoded optically in a time series [33]. The matrix multiplication is performed by combining input and weight signals and performing balanced homodyne detection. The key advantage is that the multiplication is performed passively by optical interference, so the energy costs are is 2–3 orders of magnitude smaller than for state-of-the-art CMOS circuits. However, since the multiplication is done in the photodetection, to have a high data throughput, except that two high-speed modulators are required for data and weights encoding, a high-speed balanced photodetector is indispensable, which highly increase the system costs.

In this paper, we propose and demonstrate high-throughput ONNs based on temporal computing. The core weighted summation operation is realized with the use of high-speed electro-optic modulation and low-speed balanced photodetection. The input data and weight are encoded in a time sequence separately and loaded on an optical signal via two electro-optic modulators sequentially. By precisely controlling the synchronization time of the data and weight loading, the multiplication operation is done. Followed by a balanced photodetector, the summation is conducted, thanks to the electron accumulation of the inherent electronic integrator circuit of the low-speed photodetector. Thus, the linear weighted summation operation is implemented based on temporal computing in the optical domain. With the proposed optical linear weighted summation, a fully-connected neural network and convolutional neural network are realized.

Thanks to the high-speed feature of temporal computing, a high data throughput of the ONN is experimentally demonstrated, and the weighting coefficients can be specified on demand, which enables a strong programmability of the ONN. By leveraging wavelength multiplexing technology, a scalable ONN could be created with a massive computing power and strong reconfigurability, which holds great potential for future giant AI applications.

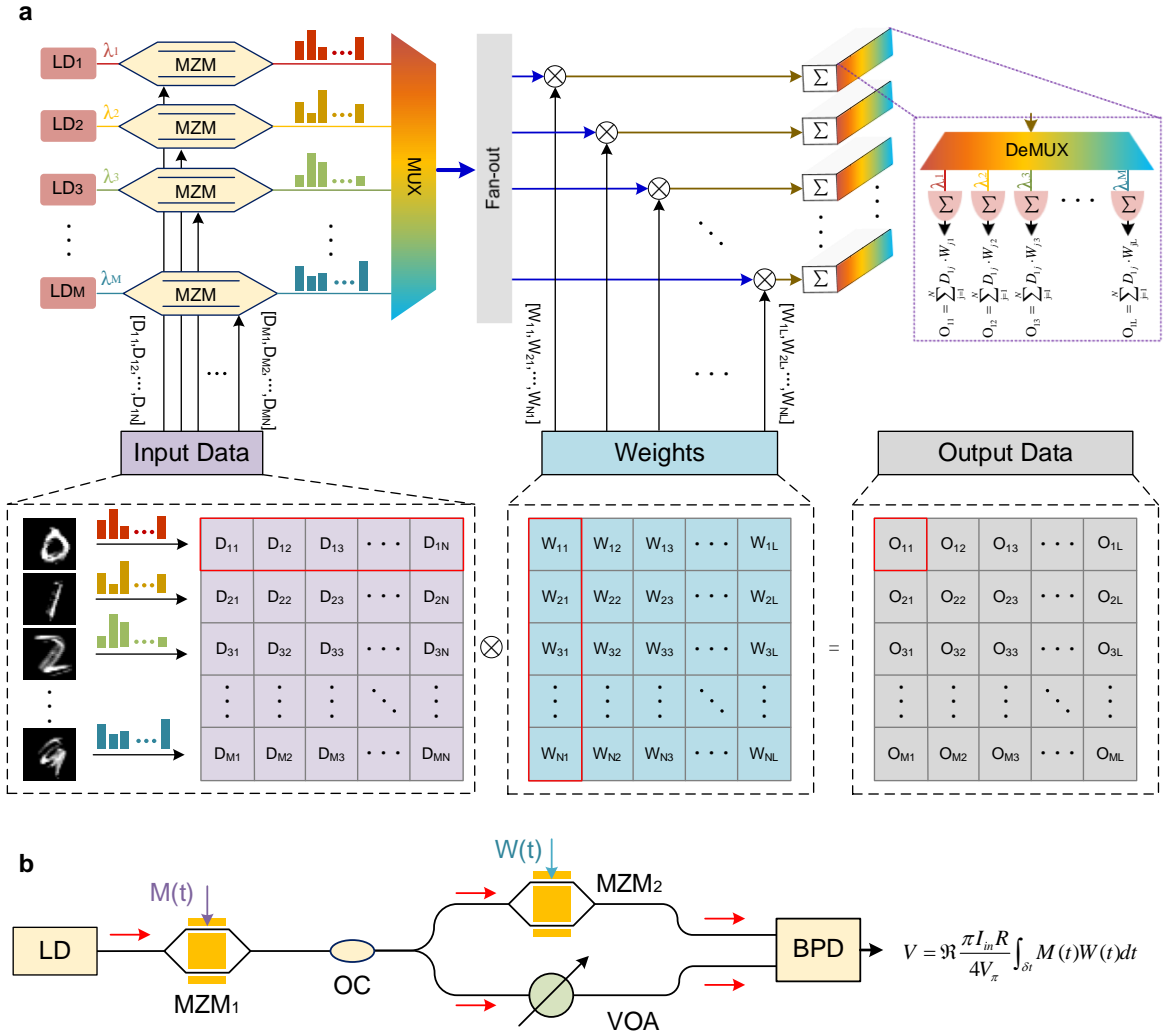

**Figure 1** (a) Schematic of the proposed high-throughput temporal optical computing system based on wavelength- and space- division multiplexing. LD, laser diode; MZM, Mach-Zehnder modulator; MUX, wavelength division multiplexer; DEMUX, wavelength division demultiplexer. (b) Principle of the proposed temporal photonic weighted summation operation unit. OC, 3-dB optical coupler; VOA, variable optical attenuator; BPD, balanced photodetector.

**Results**

Thanks to multiplexing technologies offered by optics, a high-throughput temporal optical computing system based on wavelength- and space- division multiplexing is proposed in Fig. 1(a) for the linear weighted summation operation. The input data to be processed has a $M\times N$ matrix, where $M$ different images are encoded in time as $M$ different rows with a size of $N$. With the use of $M$ high-speed electro-optic modulators, each row data sequence is loaded on different optical carrier. At the outputs of the modulators, a wavelength multiplexer is used to combine the different optical signals into a single channel. Then, the optical signals are fanned out to $L$ channels, of which each one incorporates another high-speed electro-optic modulator. The weight matrix with a size of $L\times N$ sends the weighting coefficients in each column to different modulators for multiplications. By precisely controlling the synchronization time between the data sequence and weight sequence in each channel, optical multiplication is done. At the output of each modulator in each channel, a wavelength demultiplexer is leveraged to split the different carriers into $M$ different sub-channels. In each sub-channel, a low-speed photodetector performs optical-to-electrical conversion, in which the summation is conducted, thanks to the electron accumulation of the inherent electronic integrator circuit. The system output is a $M\times L$ matrix, in which each element is a result of the linear weighted summation between a data row and a weight column.

Figure 1(b) gives the principle of the proposed temporal photonic weighted summation operation unit. The input image data to be processed is flattened into a real-valued vector and encoded as a time-domain serial electrical signal $M(t)$. With the use of a high-speed electro-optic intensity modulator MZM$_1$, the data is loaded on an optical carrier. Since the modulator MZM$_1$ is working at the null-transmission point, to compensate the waveform distortion caused by the nonlinearity of the MZM, the data signal $M(t)$ is pre-adjusted to be $M_L(t)$.

$$M_L(t) = \frac{2V_\pi}{\pi} \arcsin\left[\sqrt{M(t)}\right] \tag{1}$$

where $V_\pi$ is the half voltage of the MZM$_1$.

Thus, at the output of the MZM$_1$, the optical signal intensity can be expressed as:

$$I_{MZM1}(t) = I_{in} \sin^2\left[\frac{\pi}{2V_\pi} M_L(t)\right] = I_{in} M(t) \tag{2}$$

where $I_{in}$ is the input optical signal intensity.

Then, the modulated optical signal is divided into two paths via a 3-dB optical coupler. In the upper path, the optical signal is guided into another high-speed electro-optic intensity modulator MZM$_2$ where the weighting coefficients flattened as a vector $W(t)$ are loaded on the optical signal. By precisely controlling the synchronization time of the data and weight loading in the two MZMs, the multiplication operation between the data vector and weight vector is done. Since the MZM$_2$ is biased at the quadrature transmission point, the optical signal intensity at the output of the MZM$_2$ can be written as:

$$I_{upper}(t) = \frac{I_{MZM1}(t)}{4}\left[1 + \frac{\pi}{V_\pi}W(t)\right] \tag{3}$$

In the lower path, the optical signal undergoes a VOA, of which the intensity can be tunable.

$$I_{lower}(t) = \frac{\alpha}{2}I_{MZM1}(t) \tag{4}$$

where $\alpha$ is the attenuation coefficient of VOA. Then, a balanced photodetector (BPD) is used to collect the optical signals from the two paths and combined them for optical-to-electrical conversion. Thanks to the series connection of two PDs in the BPD, a differential photocurrent is given by:

$$i_{BPD}(t) = \Re \frac{I_{in}M(t)}{2}\left[\frac{1}{2} + \frac{\pi}{2V_\pi}W(t) - \alpha\right] \tag{5}$$

where $\Re$ is the responsivity of the BPD. When the attenuation coefficient $\alpha$ equals to 1/2, the equation can be simplified as:

$$i_{BPD}(t) = \Re \frac{\pi I_{in}}{4V_\pi}M(t)W(t) \tag{6}$$

Owing to the low-speed nature of the BPD, there is an inherent RC circuit in the commercial BPD, which can be leveraged to perform the temporal accumulation. Thus, at the output of the BPD, an electrical signal can be achieved:

$$V = \Re \frac{\pi I_{in} R}{4V_\pi}\int_{\delta t} M(t)W(t)dt \tag{7}$$

where $R$ is the parasitic resistance, $\delta t$ is the temporal accumulation window.

From Eq. 7, as can be seen, the weighted summation operation can be done with the proposed approach. By flattening the input data matrix and weight matrix into two vectors and loading them into two cascaded modulators separately in the time sequence, the vector multiplication can be realized, of which the speed is limited by the modulation bandwidth of the modulator. Usually, an available modulator in the current market is reported to have a modulation bandwidth as broad as 50 GHz [34], which enables a high-speed multiplication. In addition, a side-benefit of a strong weight reconfigurability is also enabled thanks to the high-speed feature of the modulators. To conduct the summation, a low-speed BPD is employed, which holds threefold benefits. First, thanks to its differential output, the optical multiplication can be retrieved as a differential photocurrent by canceling the common part of the two PD outputs; secondly, negative weighting coefficients are enabled; thirdly, the electric circuit in the BPD provides a straightforward solution to performing the temporal accumulation for the summation. The key limitation is that there is a temporal window on the accumulation, which is highly related to the bandwidth of the BPD. The narrower the bandwidth, the longer the temporal accumulation window.

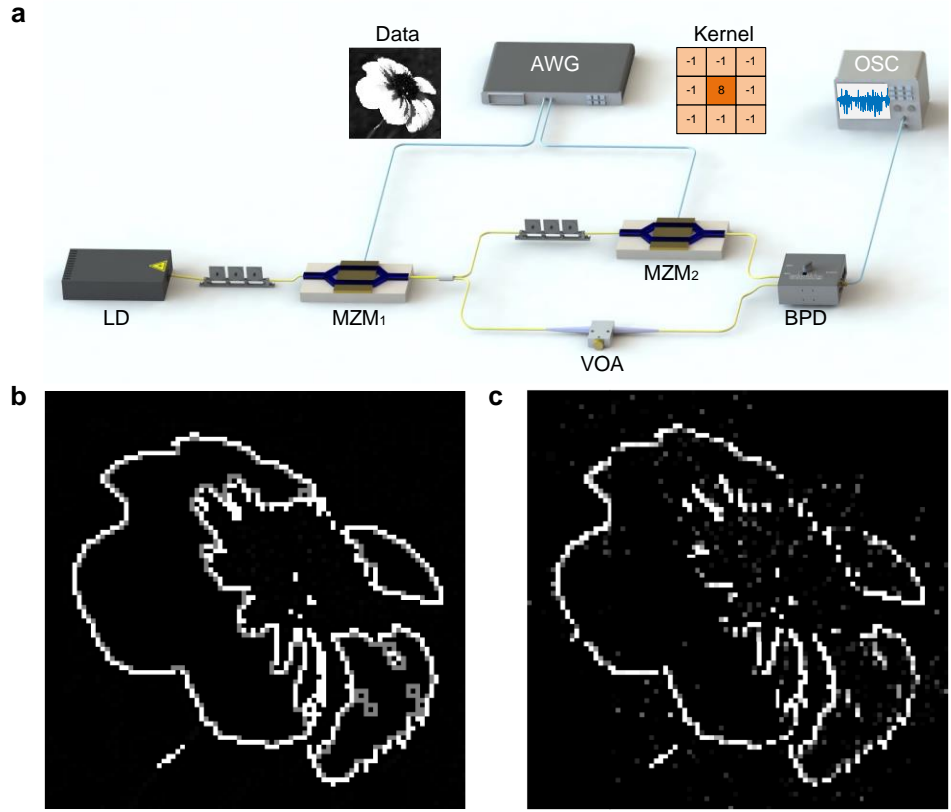

**Figure 2** (a) Experimental setup of a photonic weighted summation operation unit for image edge detection processing. LD, laser diode; MZM, Mach–Zehnder modulator; VOA, variable optical attenuator; BPD, balanced photodetector; AWG, arbitrary waveform generator; OSC, oscilloscope. (b) Theoretical result of the edge-detected image. (c) Experimental result of the edge-detected image.

Figure 2(a) shows the experimental setup of the proposed temporal photonic weighted summation operation unit. A 92×92 flower image is used as the input data, and a 3×3 Laplacian kernel is the weighting coefficients which are specified for image edge detection task. A two-channel arbitrary waveform generator (AWG) is employed to generate the data vector and kernel vector in the time sequence separately. Two high-speed MZMs are cascaded to perform the vector loading for the multiplication. A BPD with a bandwidth of 150 MHz fulfills the photodetection and temporal accumulation. The output temporal electrical signal is recorded using a real-time oscilloscope, and the edge-detected image is recovered. Figure 2(b) shows the theoretical result of the edge-detected image using the DSP algorithms. Figure 2(c) gives the experimental result of the edge-detected image using the photonic temporal computing. As can be seen, the results agree well with each other, which verifies the effectiveness of the proposed photonic weighted summation operation unit.

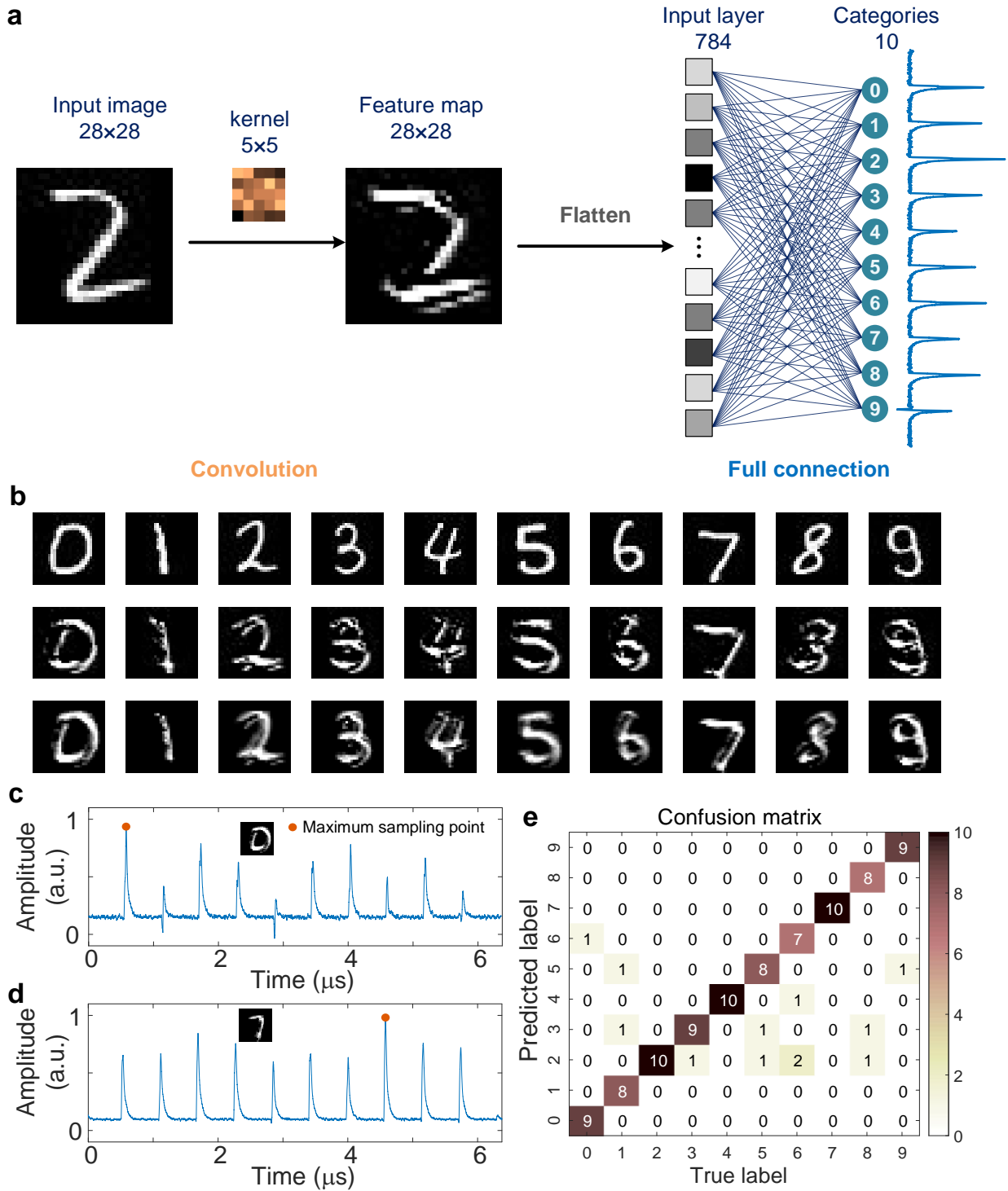

**Figure 3 An OCNN for image classification. a** Schematic of OCNN-based image classification. The OCNN consists of a convolutional layer and a fully connected layer. The linear operations in these two layers are implemented by the photonic weighted summation operation unit. **b** Theoretically calculated results and the experimental results of 10 feature maps. **c, d** Measured feature maps for fully connected layer and the corresponding temporal waveforms. **e** Confusion matrix obtained from 100 repeated experiments.

Then, the proposed photonic weighted summation operation unit is used to construct an optical convolution neural network. Convolution neural networks (CNNs), as a class of artificial neural networks, find extensive applications in visual imagery analysis, thanks to the powerful feature extraction capability [35]. Figure. 3(a) shows a typical CNN structure for classifying handwritten digit images in the MNIST dataset. The first layer is a convolutional layer, which performs feature learning; the second layer is a fully connected layer that acts as a classifier. Since in the convolution and fully connected layer, only a linear transformation is performed on the input vector through a weight matrix, these two layers can be realized with the proposed photonic weighted summation operation unit.

Figure 3(b) gives the results of the convolution layer based on the proposed photonic weighted summation operation unit. The first row is the original handwritten digit images to be processed, each of which is a 28×28 handwritten digit image. A 5×5 kernel matrix is used to extract the feature by performing the linear multiplication and accumulation operation with the input images. The operation results are reconstructed as the feature maps. The second row is the theoretically reconstructed feature maps, and the third row is experimentally reconstructed feature maps. As can be seen, the optically calculated feature maps agree well with the theoretical results, which verifies the effectiveness again of the proposed operation unit to realize the convolution layer.

With a flatten operation, each of the reconstructed feature maps is transformed to be a vector with a size of 784. To identify the handwritten digit from 0 to 9, the well-trained weight bank of the fully connected layer with a matrix size of 10×784 is divided into 10 vectors with a length of 784, and sequentially sent into the photonic weighted summation operation unit together with the feature map vectors. The operation results of the fully connected layer are recorded as a temporal waveform. Figure. 3(c) shows the experimental results of the fully connected layer when the theoretical feature map of the digit 0 is loaded. The ten peaks from left to right in the waveform correspond to the handwritten digits from 0 to 9, and the peak with the maximum value indicates the digit value. As can be seen in Fig. 3(c), the maximum sampling point corresponds to the handwritten digit 0. Figure. 3(d) shows the experimental results of the fully connected layer when the experimental feature map of the digit 7 is loaded, which agrees with the theoretically calculated one.

More numerical testing results for the classification of handwritten digits are performed. The confusion matrix for 100 testing images in Fig. 3(e) shows an accuracy of 88% for the generated predictions, in contrast to 93% for the numerical results calculated on an electrical digital computer. Due to the system instability caused by the environment perturbations and bias drifting of the MZMs, the experimental accuracy is lower than the theoretically calculated one with the identical CNN model. By leveraging on-chip integration and feedback control, the temporal computing system would have a better stability and its accuracy could be further improved. Thanks to the high-speed feature of temporal computing, a high data throughput of the ONN is experimentally demonstrated, and the weighting coefficients can be specified on demand, which enables a strong programmability of the ONN.

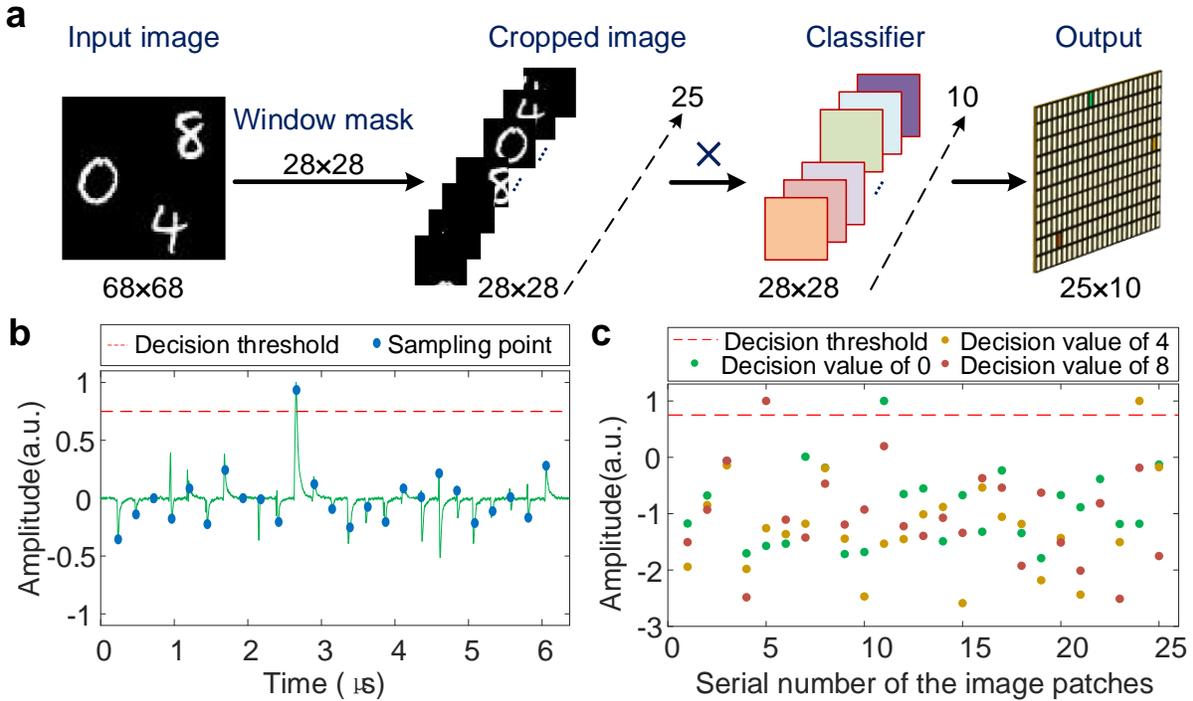

**Figure 5 Sliding-window object detection. a** Schematic of handwritten digits detection in a 68×68 image. **b** Recorded temporal waveform at the output of the BPD for the handwritten digits 0 detection. **c** Summarized weighted summation results between the 25 cropped images and three classifier matrices for digit 0, 4, and 8 detection.

**Sliding-window object detection with photonic weighted summation operation unit.** The sliding-window method, as a very classical approach in object detection, lays solid foundation for other advanced object detection algorithms [37, 38]. Figure. 5(a) shows the schematic of sliding-window multiple handwritten digits detection in an image. With a fixed 28×28 window sliding across the entire image systematically, the image with a size of 68×68 is cropped into 25 image patches in time sequence. Each cropped image is sequentially fed into a linear classifier, which consists of 10 matrixes with a size of 28×28. Each matrix is well-trained to classify each specific digit. The weighted summation operation is performed between each cropped image and each matrix. At the output of the classifier, a matrix with a size of 25×10 is generated, of which each element is the weighted summation result between each cropped image and each matrix of the classifier. The element in the output matrix indicates a decision value. By selecting the decision value that exceed the decision threshold, the corresponding handwritten digit can be detected and its position can be located in the input image according to the output matrix.

With the proposed photonic weighted summation operation unit, a sliding-window multiple digits detection is performed. Figure. 5(b) shows the recorded temporal waveform at the output of the BPD when 25 cropped images are flattened and sent into the MZM1 and the first classifier matrix for handwritten digits 0 detection is loaded on the $MZM_2$. The green line shows the weighted summation result between the 25 cropped images and the first classifier matrix. A decision threshold in the dashed red line is used to judge whether there is the digit 0 in the image patch. The sampling points that exceed the decision threshold indicate the position of the object in the original image. As can be seen, there is only one decision value larger than the threshold, which confirms the handwritten digit 0 exists. By identifying the sampling point position, the digit can be located in the $11^{st}$ cropped image, which corresponds to a specific position in the original input image.

Figure. 5(c) summarizes the weighted summation results between the 25 cropped images and the three classifier matrixes for digit 0, 4, and 8 detection. As can be seen, when the decision value is larger than the threshold, the handwritten digit can be detected. By identifying the corresponding cropped image number, the digit position in the original input image can be located. Thanks to the high throughput of the proposed photonic weighted summation approach based on temporal computing, a sliding-window multiple digit detection can be performed.

**WDM-compatible parallel photonic computing.** By leveraging wavelength multiplexing technology (WDM), a scalable ONN could be created with a massive computing power and strong reconfigurability, which holds great potential for future giant AI applications. As a proof-of-concept experimental demonstration, a WDM-compatible fully connected neural network is developed, where two wavelength-multiplexed channels are employed to realize the handwritten digit image classification in the MNIST dataset. Figure 6 shows the diagram of a parallel fully connected neural network based on WDM-compatible parallel photonic computing. Two 1×784 vectors flattened by two 28×28 images are loaded on two optical carriers via two MZMs. Then the two modulated optical signals are combined via an optical coupler and sent to another MZM for weight multiplication, where a well-trained 784×10 weight matrix is loaded. After photonic multiplication, wavelength de-multiplexing is leveraged and different wavelength are sent to different BPD for temporal accumulation. At the output of the BPD, the temporal waveform is recorded as a 1×10 vector for image classification result.

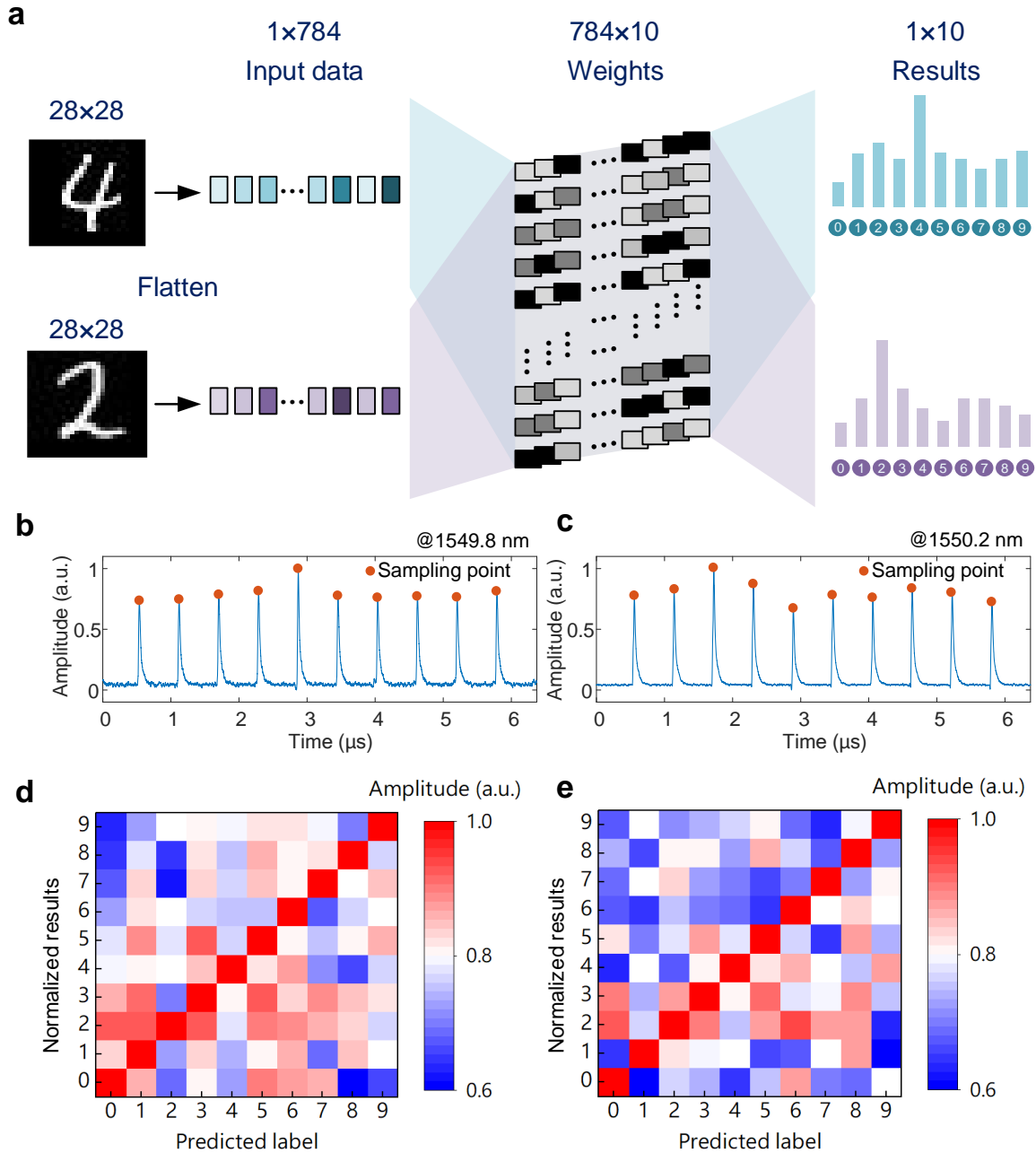

**Figure 6 Parallel fully connected neural network for image recognition task. a** Schematic of the scalable fully connected neural network based on WDM-compatible parallel photonic computing. **b, c** Measured temporal waveforms at the output of the BPDs for the handwritten digit 4 classification at 1549.8 nm and the handwritten digit 2 classification at 1550.2 nm. **d** Normalized sampling results at 1549.8 nm. **e** Normalized sampling results at 1550.2 nm.

Figure. 6(b) and (c) show the measured temporal waveforms at the output of the BPDs for the handwritten digit classification at 1549.8 nm and 1550.2 nm, respectively. As can be seen, there are 10 distinct peaks in the waveform, which represents the weight summation results. The maximum sampling point position in the measured temporal waveforms indicates the handwritten digit. Different handwritten digit images from 0 to 9 are loaded into the parallel fully connected neural network for classification. Figure 6(d) and (e) summarize the normalized sampling results for handwritten digits from 0 to 9 at the wavelength of 1549.8 nm and 1550.2 nm, which agree well with the theoretically calculated ones. Thanks to photonic temporal computing, the WDM technique can be incorporated to increase the channel number in a large scale, which would significantly enhance the computing power of the photonic computing system.

**Table 1 Comparison between the reported optical neural networks and our work.**

| Year | Ref. | Summation | Architecture | Input vector size |
|------|------|-----------|--------------|-------------------|
| 2017 | 20 | Spatial interference | MZI array | 4×1 |
| 2021 | 26 | Spatial interference | MZI array | 4×1 |
| 2023 | 29 | Spatial interference | MZI array | 4×1 |
| 2022 | 42 | Spatial interference | Cascaded MMI | 10×1 |
| 2021 | 23 | Multi-wavelength | Optical comb | 16×1 |
| 2021 | 41 | Multi-wavelength | Optical comb | 90×1 |
|      | Ours | Temporal integration | BPD | 1×784 |

## Discussion

In recent years, several different approaches have been proposed and experimentally demonstrated to realize photonic weighted summation operation for neural networks. Table 1 summarizes the previously reported work and our work in terms of summation implementation, system architecture and input vector size of the realized neural network. Optical spatial interference is a straightforward approach to realize the optical power summation. With the use of an MZI structure, optical spatial interference happens and photonic weighted summation could be done conveniently by tuning the phase difference between the two arms. However, due to its interferometric structure, stability is a key concern. Compared to discrete solutions, on-chip integrated solution is highly preferred approach by cascading numerous MZI for multiple spatial interferences. The key disadvantage is that the MZIs structure usually have a relatively large size in the scale of hundreds of micrometres, which leads to a limited number of the MZIs integrated on the same die. Consequently, it is very challenging to scale up an MZI-based ONN. From the Table, it can be seen that so far, the on-chip MZI arrays have been employed to implement a small vector size. Multi-wavelength multiplexing is another straightforward approach to realize the optical power summation. Each wavelength is an input element, and the number of the wavelength determines the input vector size. Thanks to the optical comb rapid development, a multi-wavelength optical source is possible. The key problem that the wavelength number is still limited, which inevitably leads to a limited input vector size. Consequently, the computing power is affected.

Our proposed approach leverages the temporal integration to realize the optical summation. Similar as the human brain information process based on the timing of events, temporal computing encodes the data and weight in a time sequence for multiplication and uses an integrator to fulfil the summation. In the experimental demonstration, a fully connected neural network with an input vector of $1\times784$ is realized, which holds a great potential for a massive computing power. The key advantages include high-throughput photonic weighted summation and strong scalability. The data throughput is determined by the bandwidth of the modulators and the temporal accumulation window of the BPD. With the state-of-the-art photonic transmitters and photodetectors, the weighted summation operation unit can reach at a speed of 100 GOPS [43]. By incorporating the multi-wavelength multiplexing, 100 parallel wavelength computing channels can be scaled up, of which the total effective computing speed is 10 TOPS. Furthermore, it is worth mentioning that the core of our proposed temporal computing architecture can be integrated on a single silicon photonics chip, which would lead to a reduced power consumption and much stronger scalability.

In conclusion, we propose a high-throughput photonic computing architecture based on temporal computing. The core weighted summation operation is realized with the use of high-speed electro-optic modulation and low-speed balanced photodetection. The input data and weight are encoded in a time sequence separately and loaded on an optical signal via two electro-optic modulators sequentially. By precisely controlling the synchronization time of the data and weight loading, the multiplication operation is done. Followed by a balanced photodetector, the summation is conducted, thanks to the electron accumulation of the inherent electronic integrator circuit of the low-speed photodetector. Thus, the linear weighted summation operation is implemented based on temporal computing in the optical domain. With the proposed optical linear weighted summation, a fully-connected neural network and convolutional neural network were realized. Thanks to the high-speed feature of temporal computing, a high data throughput of the ONN was experimentally demonstrated, and the weighting coefficients can be specified on demand, which enabled a strong programmability of the ONN. By leveraging wavelength multiplexing technology, a scalable ONN was created with a massive computing power and strong reconfigurability, which holds great potential for future giant AI applications.

## Methods

**Experimental setup.** The laser source (NKT Koheras BASIK) used in the experiments send an optical signal with a wavelength of 1550 nm. The image vectors and weight matrix are generated using a arbitrary waveform generator (Keysight M8196A AWG). The electro-optic modulation is realized using Fujitsu FTM7937 modulators. The BPD (Thorlabs PDB 450C) with a switchable gain and bandwidth is employed to complete the optical-to-electrical conversion. A real-time oscilloscope (Tektronix DPO75002SX) records the temporal waveform at the output of the BPD.

## References

1. Krogh, A. What are artificial neural networks? *Nat. Biotechnol*. **26**, 195–197 (2008).
2. LeCun, Y., Bengio, Y., & Hinton, G. Deep learning. *Nature* **521**(7553), 436-444 (2015).
3. Long, J., Shelhamer, E., & Darrell, T. Fully convolutional networks for semantic segmentation. In *Proceedings of the IEEE conference on computer vision and pattern recognition* 3431-3440 (2015).
4. He, K., Zhang, X., Ren, S., & Sun, J. Deep residual learning for image recognition. In *Proceedings of the IEEE conference on computer vision and pattern recognition* 770- 778 (2016).
5. Ren, S., He, K., Girshick, R., & Sun, J. Faster r-cnn: Towards real-time object detection with region proposal networks. In *Advances in neural information processing systems* 28 (2015).
6. Krizhevsky, A., Sutskever, I., & Hinton, G. E. Imagenet classification with deep convolutional neural networks. *Commun. ACM* **60**(6), 84-90 (2017).
7. Marr, B., Degnan, B., Hasler, P., & Anderson, D. Scaling energy per operation via an asynchronous pipeline. *IEEE Trans. Very Large Scale Integr. Syst.* **21**(1), 147-151 (2012).
8. Jones, N. How to stop data centres from gobbling up the world's electricity. *Nature* **561**, 163–166 (2018).
9. Dennard, R. H., Gaensslen, F. H., Yu, H. N., Rideout, V. L., Bassous, E., & LeBlanc, A. R. Design of ion-implanted MOSFET's with very small physical dimensions. *IEEE J. Solid-State Circuits* **9**(5), 256-268 (1974).

10. Tallents, G., Wagenaars, E., & Pert, G. Lithography at EUV wavelengths. *Nat. Photonics* **4**(12), 809-811 (2010).

11. Caulfield, H. J. & Dolev, S. Why future supercomputing requires optics. *Nat. Photonics* **4**(5), 261–263 (2010).

12. Solli, D. R., & Jalali, B. Analog optical computing. *Nat. Photonics* **9**(11), 704-706 (2015).

13. Shastri, B. J., Tait, A. N., Ferreira de Lima, T., Pernice, W. H., Bhaskaran, H., Wright, C. D., & Prucnal, P. R. Photonics for artificial intelligence and neuromorphic computing. *Nat. Photonics* **15**(2), 102-114 (2021).

14. Zhou, H. et al. Photonic matrix multiplication lights up photonic accelerator and beyond. *Light Sci. Appl.* **11**(1), 1-21 (2022).

15. Zuo, Y., et al. All-optical neural network with nonlinear activation functions. *Optica* **6**(9), 1132-1137 (2019).

16. Miscuglio, M. et al. Massively parallel amplitude-only Fourier neural network. *Optica* **7**(12), 1812-1819 (2020).

17. Lin, X. et al. All-optical machine learning using diffractive deep neural networks. *Science* **361**, 1004–1008 (2018).

18. Zhou, T. et al. Large-scale neuromorphic optoelectronic computing with a reconfigurable diffractive processing unit. *Nat. Photonics* **15**(5), 367-373 (2021).

19. Vandoorne, K. et al. Experimental demonstration of reservoir computing on a silicon photonics chip. *Nat. Commun.* **5**, 3541 (2014).

20. Shen, Y., et al. Deep learning with coherent nanophotonic circuits. *Nat. Photonics* **11**(7), 441-446 (2017).

21. Feldmann, J., Youngblood, N., Wright, C. D., Bhaskaran, H., & Pernice, W. H. All-optical spiking neurosynaptic networks with self-learning capabilities. *Nature* **569**(7755), 208-214 (2019).

22. Huang, C. et al. Demonstration of scalable microring weight bank control for large-scale photonic integrated circuits. *APL Photonics* **5**(4), 040803 (2020).

23. Feldmann, J., et al. Parallel convolutional processing using an integrated photonic tensor core. *Nature* **589**(7840), 52-58 (2021).

24. Huang, C. et al. A silicon photonic–electronic neural network for fibre nonlinearity compensation. *Nat. Electron.* **4**(11), 837-844 (2021).

25. Xu, S., et al. Optical coherent dot-product chip for sophisticated deep learning regression. *Light Sci. Appl.* **10**(1), 221 (2021).

26. Zhang, H., et al. An optical neural chip for implementing complex-valued neural network. *Nat. Commun.* **12**(1), 457 (2021).

27. Wu, C., Yu, H., Lee, S., Peng, R., Takeuchi, I., & Li, M. Programmable phase-change metasurfaces on waveguides for multimode photonic convolutional neural network. *Nat. Commun.* **12**(1), 96 (2021).

28. Xu, S., Wang, J., Yi, S., & Zou, W. High-order tensor flow processing using integrated photonic circuits. *Nat. Commun.* **13**(1), 7970 (2022).

29. Huang, Y., et al. Easily scalable photonic tensor core based on tunable units with single internal phase shifters. *Laser & Photonics Rev.* 2300001 (2023).

30. Tait, A. N. et al. Neuromorphic photonic networks using silicon photonic weight banks. *Sci. Rep.* **7**, 7430 (2017).

31. Everson, L. R., Liu, M., Pande, N., & Kim, C. H. A 104.8 TOPS/W one-shot time-based neuromorphic chip employing dynamic threshold error correction in 65nm. *In 2018 IEEE Asian Solid-State Circuits Conference (A-SSCC)* 273-276 (2018).

32. Sludds, A., et al. Delocalized photonic deep learning on the internet's edge. *Science* **378**(6617), 270-276 (2022).

33. Hamerly, R., Bernstein, L., Sludds, A., Soljačić, M. & Englund, D. Large-scale optical neural networks based on photoelectric multiplication. *Phys. Rev. X* **9**, 021032 (2019).

34. Li, M., Wang, L., Li, X., Xiao, X., & Yu, S. Silicon intensity Mach–Zehnder modulator for single lane 100 Gb/s applications. *Photonics Res.* **6**(2), 109–116 (2018)

35. Li, Z., Liu, F., Yang, W., Peng, S., & Zhou, J. A survey of convolutional neural networks: analysis, applications, and prospects. *IEEE Trans. Neural Netw. Learn. Sys.* (2021).

36. Xu, X., et al. Photonic perceptron based on a kerr microcomb for high-speed, scalable, optical neural networks. *Laser Photonics Rev.* **14**(10), 2000070 (2020).

37. LeCun, Y., Bengio, Y. Convolutional networks for images, speech, and time series. *The handbook of brain theory and neural networks* **3361**(10) (1995).

38. Wang, X., Han, T. X., & Yan, S. An HOG-LBP human detector with partial occlusion handling. In *2009 IEEE 12th International Conference on Computer Vision* 32-39 (2009).

39. Felzenszwalb, P. F., Girshick, R. B., & McAllester, D. Cascade object detection with deformable part models. In *2010 IEEE Computer Society Conference on Computer Vision and Pattern Recognition* 2241-2248 (2010).

40. Lawrence S., Giles C. L., Tsoi A. C., et al. Face recognition: A convolutional neural-network approach. *IEEE trans. neural netw.* **8**(1) 98-113 (1997).

41. Xu, X. et al. 11 TOPS photonic convolutional accelerator for optical neural networks. *Nature* **589**(7840), 44-51 (2021).

42. Zhu, H. H., et al. Space-efficient optical computing with an integrated chip diffractive neural network. *Nat. Commun.* **13**(1), 1-9 (2022).

43. He, M., et al. High-performance hybrid silicon and lithium niobate Mach–Zehnder modulators for 100 Gbit s−1 and beyond. *Nat. Photonics* **13**(5), 359-364 (2019).

## Supplementary information

Supplementary Information is linked to the online version of the paper.

## Author contributions

S.Z., J.Z. and W.Z. conceived the idea. J.Z. and S.Z. performed the experiments. S.Z. and J.Z. analyzed the data. S.Z. and J.Z. contributed to the discussion of experimental results. W.Z. supervised and coordinated all the work. All authors co-wrote the manuscript.

## Acknowledgments

This work was supported by the National Key R&D Program of China (2019YFB2203303), National Natural Science Foundation of China (NSFC) (62105028 and 62071042).

## Data Availability

The data that support the findings of this study are available from the corresponding authors on reasonable request.

## Competing interests

The authors declare no competing interests.